\documentclass[preprint,preprintnumbers,amsmath,amssymb]{revtex4}
\usepackage{graphicx}
\usepackage{dcolumn}
\usepackage{bm}

\begin{document}

\title[Secondly Revised Manuscript for CTP]{Combined Influence of Off-diagonal System Tensors
 and Potential Valley Returning of the Optimal Path}

\author{Chun-Yang Wang}\thanks{Corresponding author. Electronic mail: wchy@mail.bnu.edu.cn}

\address{Department of Physics, Shandong Provincial Key
Laboratory of Laser Polarization and Information Technology, Qufu
Normal University, Qufu 273165, China}

\begin{abstract}
The two-dimensional barrier passage is studied in the framework of
Langevin statistical reactive dynamics. The optimal incident angle
for a particle diffusing in the dissipative non-orthogonal
environment with various strengths of coupling between the two
degrees of freedom is systematically calculated. The optimal
diffusion path of the particle in a non-Ohmic damping system is
revealed to have a probability to return to the potential valley
under the combined influence of the off-diagonal system tensors.
\end{abstract}

 \maketitle

\section{Introduction}

The problem of escaping from a metastable state potential is of
ubiquitous interest in almost all scientific areas in particular the
study of nuclear reactions. The related reactive system is usually
modeled within the framework of one-dimensional (1D) standard
Brownian motion \cite{sbm1,sbm2,sbm3,hof}. However, since it leaves
out the correlation between different degrees of freedom while many
processes obviously involve more than one degree of freedom, the
classical 1D model fails to describe satisfactorily the dynamical
evolving of a real reactive process. Basing on these considerations
we have recently generalized the 1D model to the two-dimensional
(2D) case by analyzing a set of coupled generalized Langevin
eqnarry\cite{Chyw,Chyw2}.

It has been shown that the diffusion in a 2D potential energy
surface (PES) includes nutritious useful information of the reactive
dynamics such as there exists an optimal incident angle (or an
optimal path) for the particle to obtain its maximum probability to
surmount the PES barrier. This provides a convenient way to
understand many stochastic dynamical processes such as the fusion of
massive nuclei and even the synthesis of super-heavy elements
because one can easily estimate the reactive probability of a
particle by tracing its footprint along the optimal path.

Moreover, It is also revealed that the non-orthogonality of the PES
and the off-diagonal system parameters is very important in
determining whether a reactive processes can be easily accomplished
or not. Although this was mentioned in some previous
studies\cite{off-dia1,off-dia2}, less effort has been made to give a
thorough investigation. In particular, no research has based
directly on the optimal incident angle as far as we have known.
Therefore it is very meaningful to seek for more detailed
information on this subject.

In this paper, motivated by the interest of better understanding the
2D reactive dynamics, we present a relatively systematic study of
the optimal incident angle which enables the particle to surmount
the barrier with maximum passing probability. Firstly, in Sec.
\ref{sec2} a large number of Langevin calculations are performed in
the Ohmic damping case by simultaneously varying the off-diagonal
term of the system tensors. Secondly, in Sec. \ref{sec3}, influence
of the non-orthogonality of the system tensors on the non-Ohmic
damping diffusion process is discussed where a startling potential
valley returning behavior of the optimal path is witnessed. Sec.
\ref{sec4} is a summary of our conclusion in which also the implicit
application of this work is discussed.

\section{Combined Influence of Off-diagonal System Tensors\label{sec2}}

In brief, we begin with the optimal incident angle defined in our
previous study by tracing the minimum value of critical initial
velocity $v^{c}_{_{0}}$ in a 2D Ohmic damping barrier surmounting
process\cite{Chyw}. It reads
\begin{eqnarray}
\phi_{m}=\textrm{arctan}\left(\frac{m_{12}(\beta_{22}a+\omega_{22})-m_{22}(\beta_{12}a+\omega_{12})
}{m_{11}\textsl{F}_{1}(a)+m_{12}\textsl{F}_{2}(a)}\right)\label{eq1}
\end{eqnarray}
where $\textsl{F}_{j}(a)$s, $(j=1,2)$ are the related functions
derived in the process of Laplace solving the coupled Langevin
eqnarray
$m_{ij}\ddot{x}_{j}(t)+\beta_{ij}\dot{x}_{j}(t)+\omega_{ij}x_{j}(t)=\xi_{i}(t)$.
Expressions of them are
\begin{eqnarray}
\textsl{F}_{1}(a)&=&m_{22}a^{2} +\beta_{22}a+\omega_{22},\nonumber\\
\textsl{F}_{2}(a)&=&-m_{12}a^{2}-\beta_{12}a-\omega_{12},\label{eq2}
\end{eqnarray}
with $a$ the largest analytical root of
\begin{eqnarray}
&&(\textrm{det}m)s^{4}+(m_{11}\beta_{22}+m_{22}\beta_{11}-2m_{12}\beta_{12})s^{3}
\nonumber\\ && \ \ \ \
+(\textrm{det}\beta+m_{11}\omega_{22}+m_{22}\omega_{11}-2m_{12}\omega_{12})s^{2}\nonumber\\&&
\ \ \ \
+(\beta_{11}\omega_{22}+\beta_{22}\omega_{11}-2\beta_{12}\omega_{12})s+\textrm{det}\omega=0\label{eq3}
\end{eqnarray}
the symbols with subscripts such as $m_{12}$ are the components of
inertia ($m_{ij}$), friction ($\beta_{ij}$) and potential-curvature
($\omega_{ij}$) tensors respectively, ``det" denotes the determinant
of each tensor. In the Ohmic damping case, all the system tensors in
the Langevin eqnarray can be considered as invariable constants and
correlations of the two components of the random force $\xi_{i}(t)$
obey the fluctuation-dissipation theorem $
\langle\xi_{i}(t)\xi_{j}(t')\rangle=k_{_{B}}Tm_{ik}^{-1}\beta_{kj}\delta(t-t')$,
where $k_B$ is the Blotzmann constant and $T$ the temperature.

\begin{figure}
\centering
\includegraphics[scale=0.8]{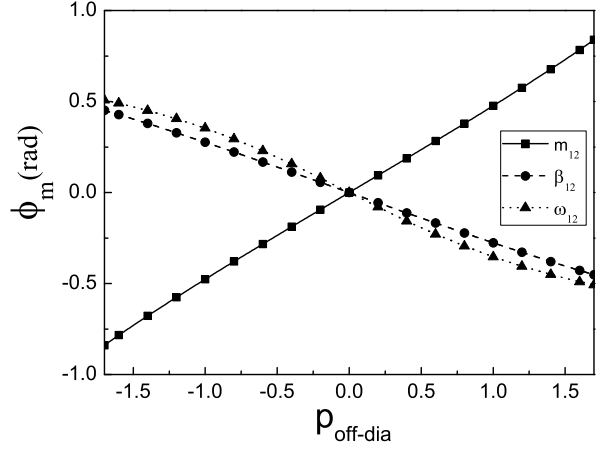}
\centering \caption{Optimal incident angle as a function of various
off-diagonal parameters ($\textsf{p}_{\textsf{off-dia}}$). When one
is varying the other two remains zero. Other parameters used here
are: $m_{11}=1.5$, $m_{22}=2.0$, $\beta_{11}=1.8$, $\beta_{22}=1.2$,
$\omega_{11}=-2.0$, $\omega_{22}=1.5$.\label{fig1}}
\end{figure}

\begin{figure}
\centering
\includegraphics[scale=0.6]{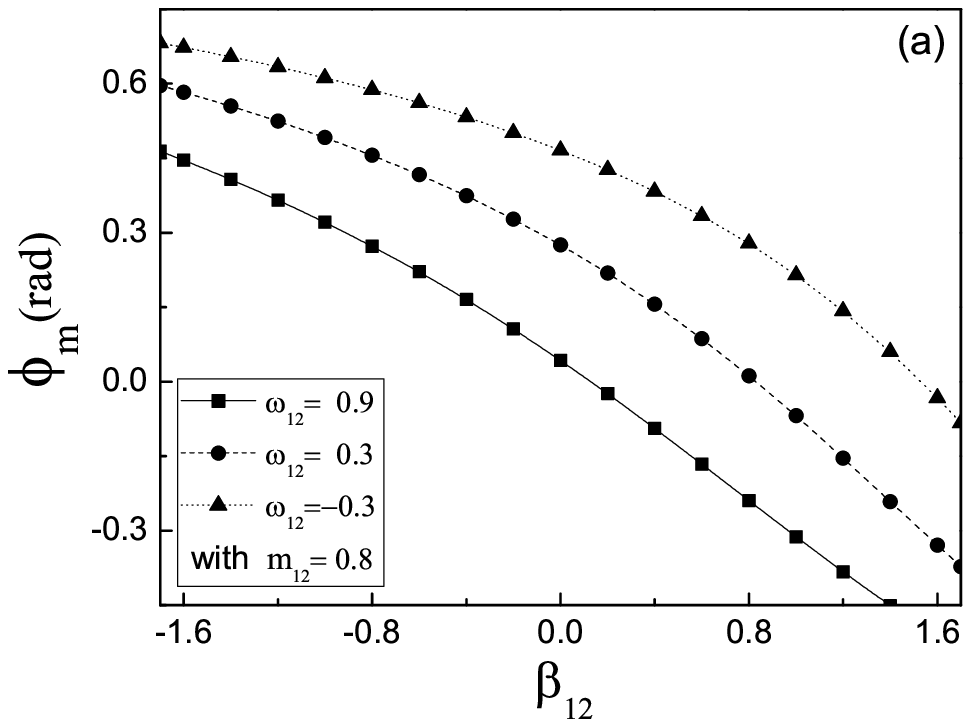}
\includegraphics[scale=0.6]{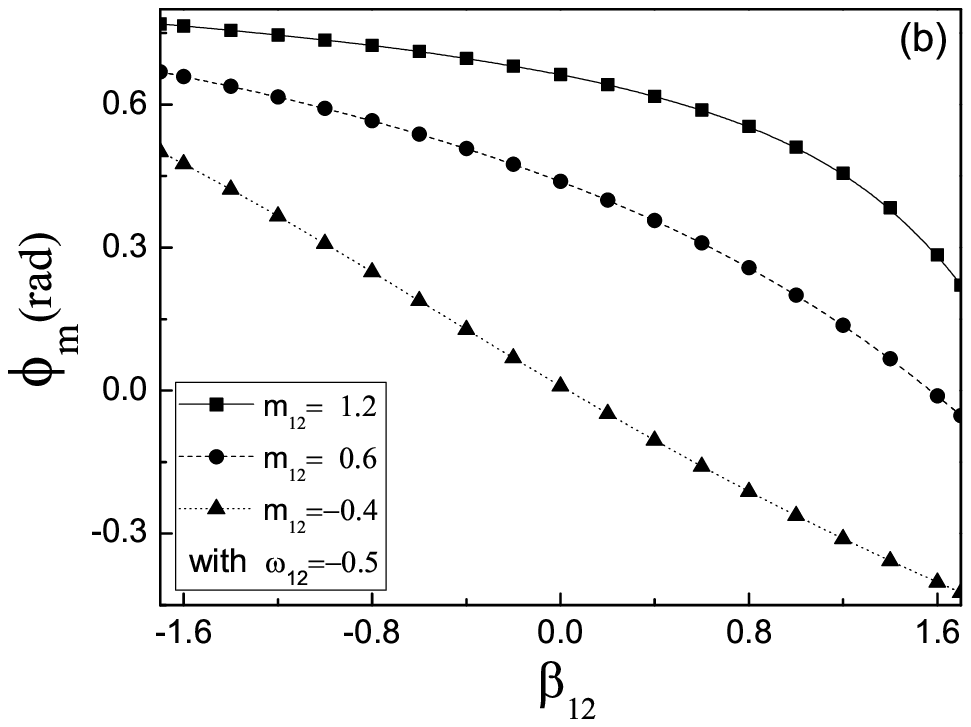}
\includegraphics[scale=0.6]{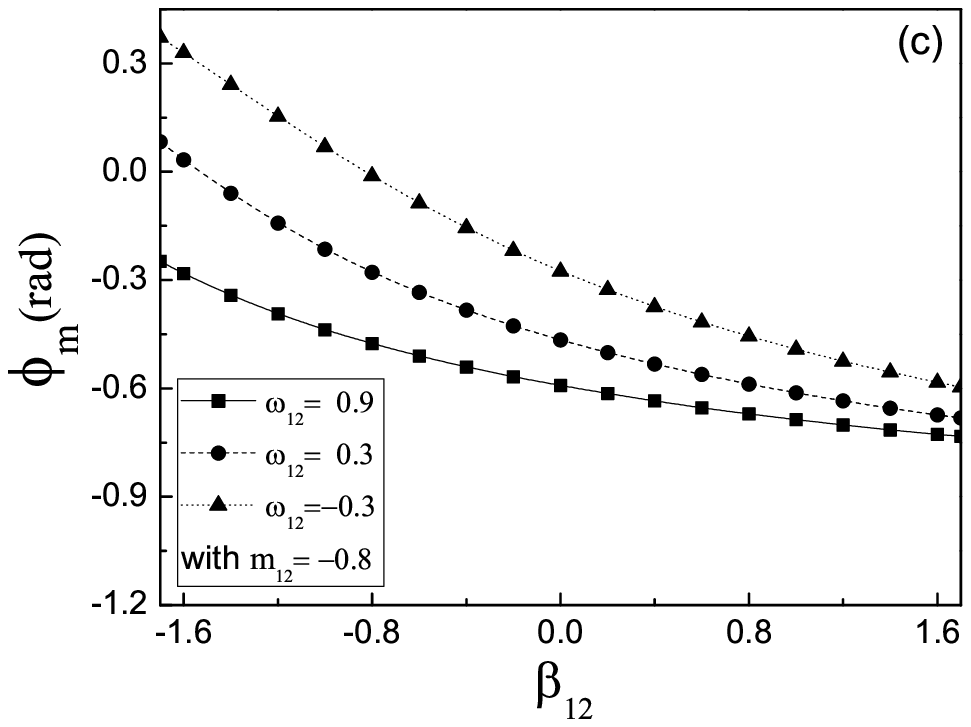}
\includegraphics[scale=0.6]{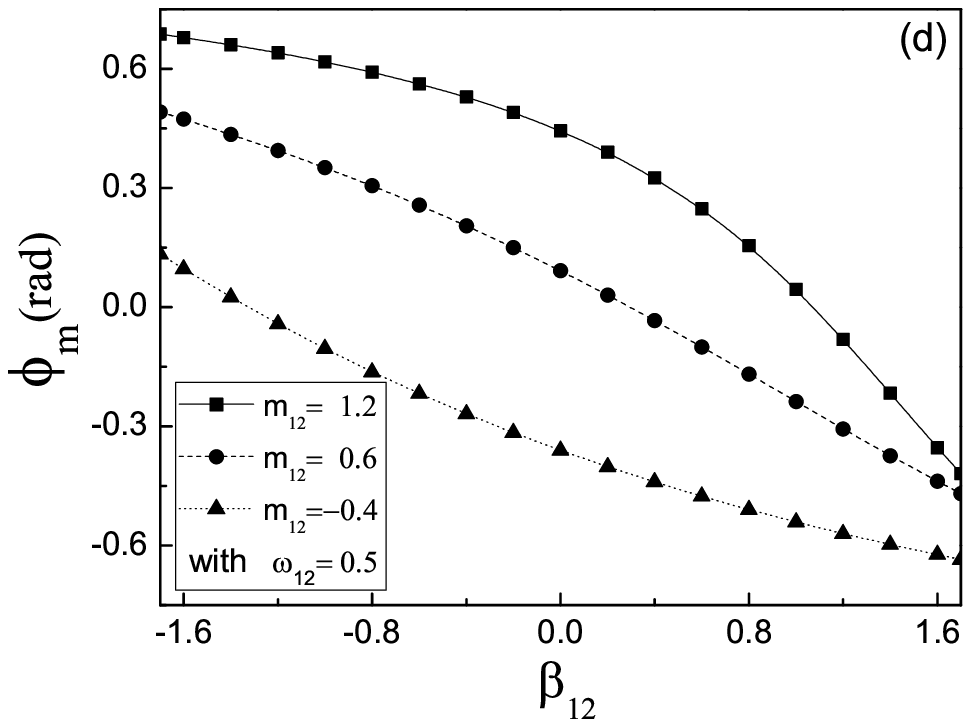}
\centering \caption{Optimal incident angle as a function of
$\beta_{12}$ for various $\omega_{12}$ (or $m_{12}$). Identical
system parameters are used as those given in Fig. \ref{fig1}.
\label{fig2}}
\end{figure}

Noticing, Eq. (\ref{eq1}) implicitly contains the off-diagonal term
of the system tensors. This implies the non-orthogonality of the
system tensors is an important factor in determining the reactive
probability of a particle. A combined control of them on the optimal
incident angle can be expected. The primary purpose of this work is
then to obtain some comprehension on this phenomenon.

   In the calculations here and following, we rescale all the variables
so that the dimensionless unit is used. In Fig. \ref{fig1}, the
optimal incident angle is plotted as a function of various
off-diagonal parameters ($\textsf{p}_{\textsf{off-dia}}$). From
which we can see that the optimal incident angle varies almost
linearly from positive to negative as the increasing of
$\omega_{12}$ (or $\beta_{12}$) when there is no influence of other
off-diagonal ones, while the effect of $m_{12}$ is completely on the
contrary. Supposing the diffusion takes place in a 2D $x_{1}-x_{2}$
PES and $x_{1}$ is the potential valley direction, it implies that
the non-orthogonality of the potential-curvature (or the friction)
makes the average diffusion path of the particle to turn toward the
negative $x_{2}$ axis, while the asymmetry of the inertia has a
opposite affection. This reveals, in a certain reactive process, the
optimal incident angle may to some extent be controlled by the
varying of the non-orthogonality of the system tensors. The
co-operation of the three off-diagonal system tensors may lead to a
relatively ideal dynamical reactive path for the particle to
surmount the barrier.

   In order to get more detailed information, we made a thorough analysis
about the influence of the non-orthogonality on the reactive
dynamics by considering simultaneously the varying of all the three
off-diagonal components. In Fig. \ref{fig2}, the optimal incident
angle is plotted as a function of $\beta_{12}$ for various
$\omega_{12}$ (or $m_{12}$) at certain $m_{12}$ (or $\omega_{12}$).
A common character is found as that the optimal incident angle
decreases almost linearly as the increasing of $\beta_{12}$. However
the decaying rate is discordance in each subgraph. An important
conclusion can be made by comparing each curve in Fig. \ref{fig2} as
that for a reactive system with definite off-diagonal friction
tensor the non-orthogonality of the potential-curvature or symmetry
of the inertia could help the diffusing particle to obtain a big
probability to pass the barrier. This is of directive significance
in the simulation or experimental operation of many reactive
processes such as the fusion of massive nuclei because accordingly
one could try to select symmetric collision and non-orthogonal
approximation of the potential to obtain an ultimate fusion
probability.

\begin{figure}
\centering
\includegraphics[scale=0.6]{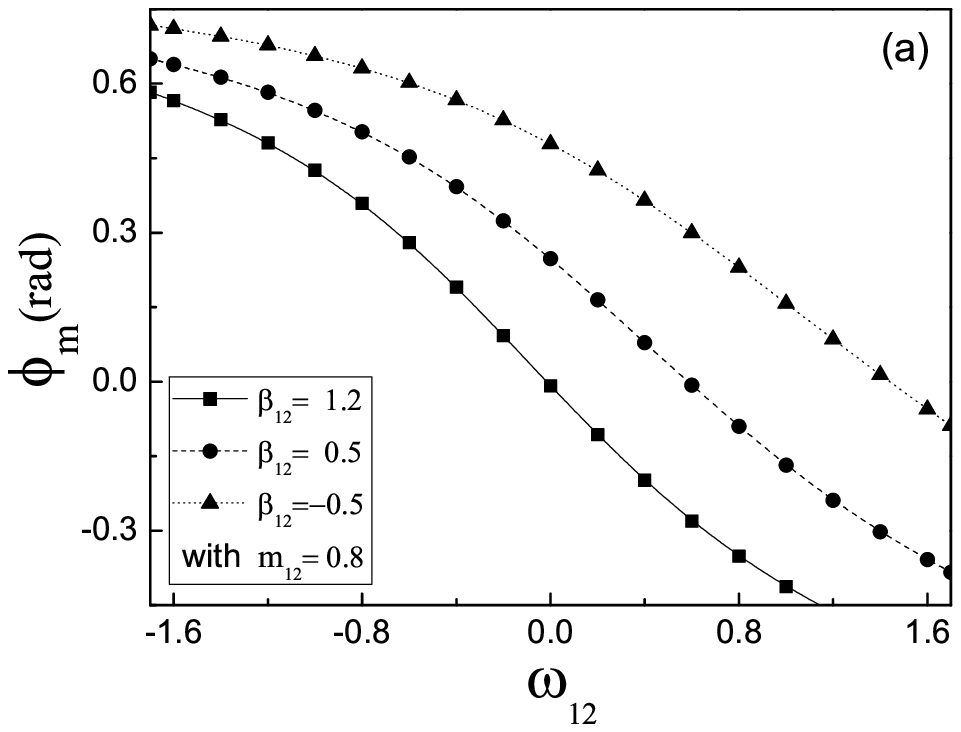}
\includegraphics[scale=0.6]{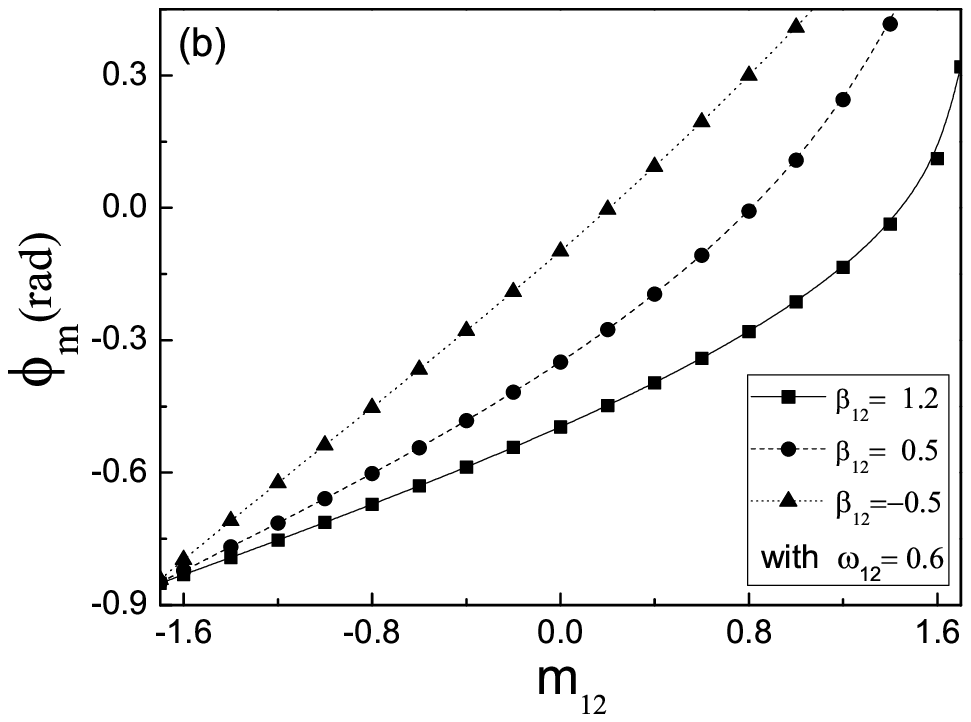}
\includegraphics[scale=0.6]{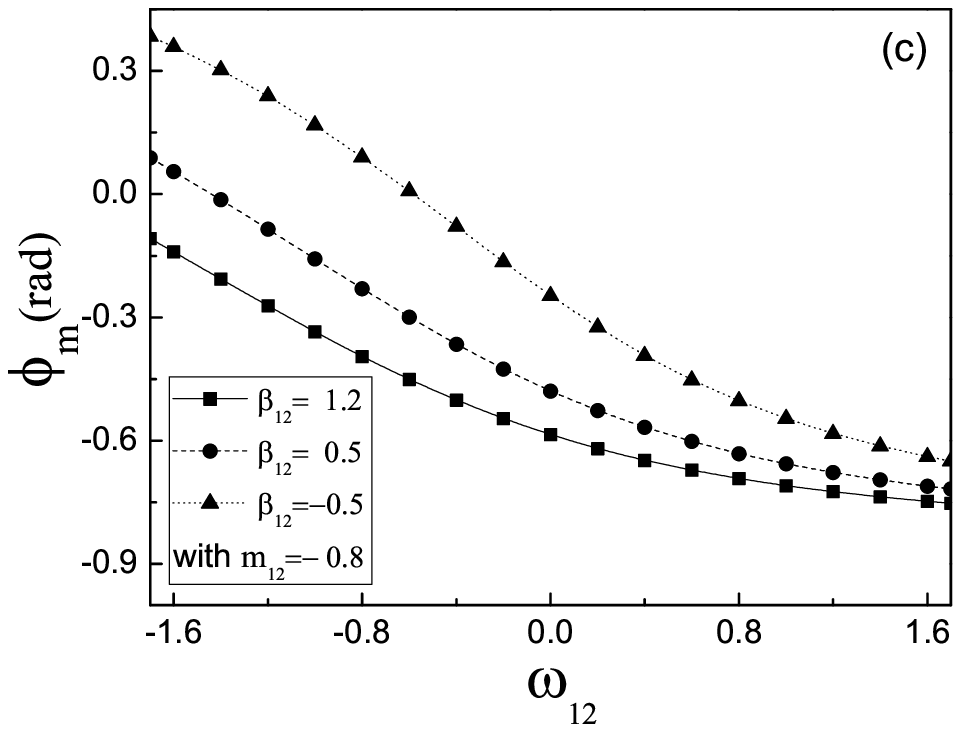}
\includegraphics[scale=0.6]{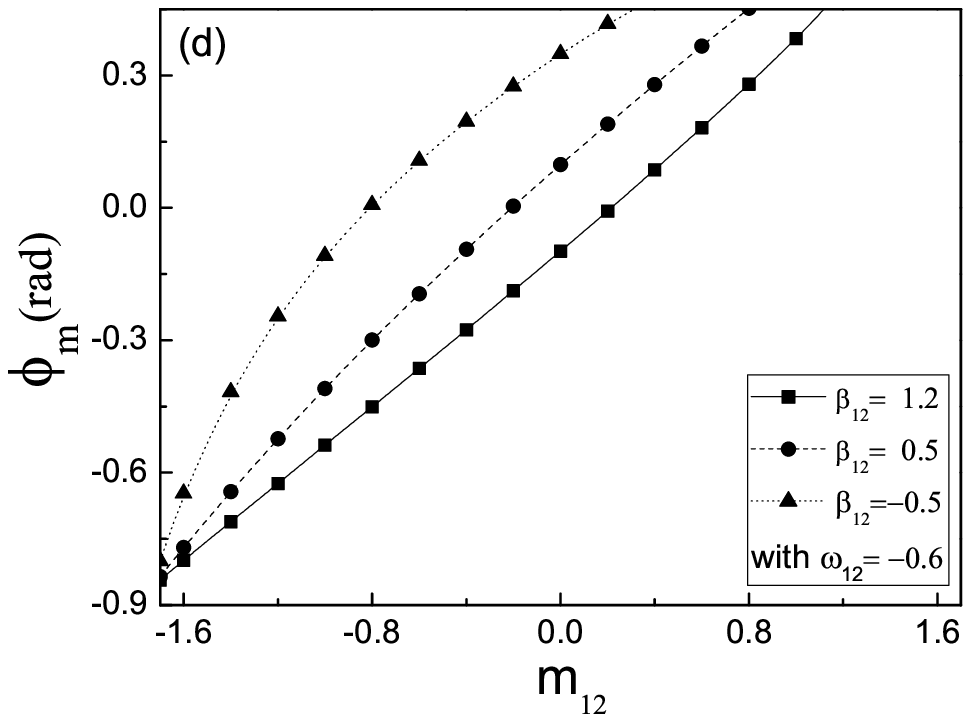}
\caption{Optimal incident angle as a function of $\omega_{12}$ (or
$m_{12}$) for various $\beta_{12}$. Identical system parameters are
used as those given in Fig. \ref{fig1}.\label{fig3}}
\end{figure}

For comparison, in Fig. \ref{fig3}, the optimal incident angle is
plotted as a function of $\omega_{12}$ (or $m_{12}$) for various
$\beta_{12}$. From which we can see, the optimal incident angle
decreases (or increases) almost linearly with the increasing of
$\omega_{12}$ (or $m_{12}$) at various rates. But noticing in
subgraphs Fig. \ref{fig3} (a) and (c) a large value of $\beta_{12}$
makes it easy for the optimal incident angle to reach zero (the
potential valley direction) while in subgraphs Fig. \ref{fig3} (b)
and (d) a small $\beta_{12}$ is appreciated. This reveals, given the
potential and inertia asymmetry is definite, the influence of the
non-orthogonality of friction on the diffusion process is relatively
complicated. This is comprehensible in the fusion process of massive
nuclei. From the view point of diffusion induces fusion reaction,
the friction of a reactive system relies mostly on the coupling
between the system and the bath environment. In the 2D case we
concerned, the strength of friction is also restricted by the
coupling between two degrees of freedom. Thus results in for the
diffusing particle a time-dependent or coordinate-dependent
environment which makes the barrier surmounting process complicate.

\section{Potential valley returning of the optimal path\label{sec3}}

\begin{figure}
\centering
\includegraphics[scale=0.8]{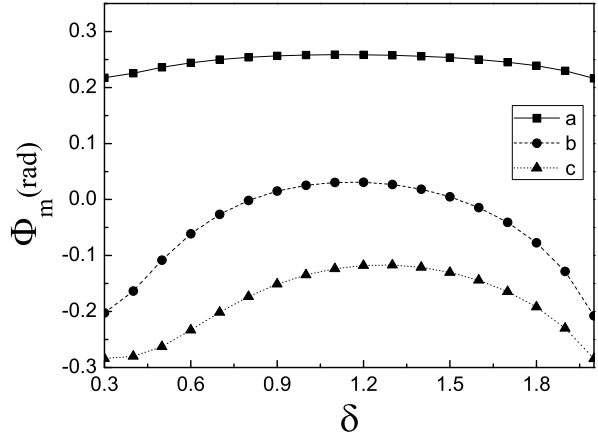}
\caption{Non-Ohmic optimal incident angle as a function of the power
exponents $\delta$ for various effective frictions. The system
parameters for each curve are (a) $\gamma_{12}=0.8$,
$\gamma_{11}=1.8$, $\gamma_{22}=1.2$; (b) $\gamma_{12}=2.0$,
$\gamma_{11}=2.5$, $\gamma_{22}=2.2$; (c) $\gamma_{12}=4.0$,
$\gamma_{11}=4.8$, $\gamma_{22}=4.2$, with $\omega_{12}=-0.5$ and
$m_{12}=0.6$ except those diagonal components of each tensor as
those given in Fig. \ref{fig1}. \label{fig4}}
\end{figure}

In order to have a deep comprehension on this phenomenon, we
consider a type of time-dependent system friction resulted from the
non-Ohmic power spectral density
$J_{ij}(\omega)=\gamma_{ij}(\omega/\omega_{r})^{\delta}$\cite{nonOhmic1,nonOhmic2,nonOhmic3,JCP-Bao},
 where $\delta$ is the power exponent taking values between 0 and 2
 in which $\delta=1.0$ corresponds to the Ohmic damping case discussed above. $\gamma_{ij}$ is the
symmetrical friction constant tensor, and $\omega_{r}$ denotes a
reference frequency induced in order to ensure the components of
$\gamma_{ij}$ to have the dimension of a viscosity at any $\delta$.
This kind of non-Ohmic damping can describe a large group of
anomalous diffusions \cite{nsp1,nsp2,nsp3}.

The leading part in the following discussion is still based on the
optimal incident angle. In the non-Ohmic damping case it can be
obtained by a simple generalization from Eqs.(\ref{eq1})-(\ref{eq3})
with all the components of friction $\beta_{ij}$ $(i,j=1,2)$
replaced by the corresponding Laplacian transformation
$\hat{\beta}_{ij}[s]=\tilde{\gamma}_{ij}s^{\delta-1}$\cite{Chyw2}.
Where
$\tilde{\gamma}_{ij}=\gamma_{ij}\omega_{r}^{1-\delta}\textrm{sin}^{-1}(\delta\pi/2)$
is the effective friction with each component of $\gamma_{ij}$ being
supposed to be a constant independent of $\delta$. Thus the optimal
incident angle in the non-Ohmic case reads
\begin{eqnarray}
\Phi_{m}=\textrm{arctan}\left(\frac{m_{12}(\hat{\beta}_{22}[s]\textrm{a}
+\omega_{22})-m_{22}(\hat{\beta}_{12}[s]\textrm{a}+\omega_{12})
}{m_{11}\textsl{F}_{1}(a)+m_{12}\textsl{F}_{2}(a)}\right),\label{eq,oa}
\end{eqnarray}
implicitly containing the power exponent $\delta$.

In Fig. \ref{fig4}, the non-Ohmic optimal incident angle $\Phi_{m}$
is plotted as a function of exponent $\delta$ at various strengths
of effective frictions. In which it is revealed that the non-Ohmic
optimal incident angle evolves as a non-monotonic function of the
exponent $\delta$. Given the effective friction is relatively strong
(seen line (b) and (c) in Fig. \ref{fig4} for example ) the optimal
incident angle tends to approach zero as $\delta$ is varying from
non-Ohmic region to Ohmic case ($\delta=1.0$). This is an un-trivial
behavior because $\Phi_{m}=0$ represents the potential valley
direction. So an amazing prediction can be made as that the optimal
diffusion path will in some case return to the potential valley as
is always expected in the 1D model. It is also distinguished from
the 2D Ohmic damping case where the optimal diffusion path is
usually considered deviating from the potential valley
direction\cite{Chyw}. This can be understood from the view point of
combined influence of the non-orthogonality of system tensors and
the coupling resulted 2D non-Ohmic damping environment. It is their
combination who lead the returning of 2D non-Ohmic optimal diffusion
path to the potential valley.

\begin{figure}
\centering
\includegraphics[scale=0.8]{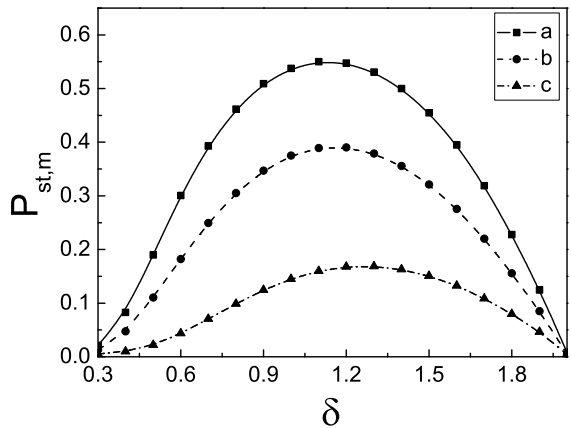}
\caption{The maximum stationary passing probability as a function of
the power exponents $\delta$ for various effective frictions.
Identical system parameters are used as those in Fig. \ref{fig4}.
\label{fig5}}
\end{figure}

This can also be understood by investigating the stationary barrier
passing probability of the particle which is usually known as a
supplemented complement error function
$P_{st}=\frac{1}{2}\textrm{Erfc}\left[-\langle
x_{1}(t)\rangle/\sqrt{2}\sigma_{x_{1}}(t)\right]$ of the reactive
degree of freedom denoted as $x_{1}$ here \cite{Chyw2}. In fig.
\ref{fig5} we plot the maximum value of stationary barrier passing
probability $P_{st,m}$ (it can be got by tracing the particle along
the optimal path) as a function of the power exponents $\delta$ for
various effective frictions. From which we can see, at most cases of
non-Ohmic damping ($\delta\neq1.0$) the maximum stationary barrier
passing probability is smaller than the Ohmic damping case
($\delta=1.0$) except for a narrow range of $1.0<\delta<1.4$. This
is because different values of power exponent $\delta$ results in
different values of critical initial velocity. For example, one can
find $v^{c}_{_{0}}|_{\delta=1.8}\cong2.6417>v^{c}_{_{0}}
|_{\delta=0.6}\cong2.3477>v^{c}_{_{0}}|_{\delta=1.0}\cong1.8426$,
while
$v^{c}_{_{0}}|_{\delta=1.1}\cong1.8251<v^{c}_{_{0}}|_{\delta=1.0}\cong1.8426$,
calculated by using of the system parameters presented in Fig.
\ref{fig4}(a) and incident angle $\Phi=0.258\textmd{rad}$. Thus it
results in the non-monotonic behavior of the maximum stationary
barrier passing probability $P_{st,m}$ and then the potential valley
returning of the optimal path.

Although it is not the primary purpose of this work, determination
of the fusion probability is very important in the study of fusion
dynamics. The potential valley returning behavior of the optimal
path revealed in present study will provide very useful information
for the experimental studying of a real reactive process. This is
because one can accordingly select the most appropriated combination
of the non-orthogonality of system tensors or try to make some
possible adjustment on the dissipative environment to obtain a big
fusion probability. For example, symmetric collision and
non-orthogonal approximation of the potential in Ohmic damping
environment and suitable friction strength in non-Ohmic case are
appropriated as revealed in our study.

\section{Summary\label{sec4}}

In conclusion of this paper, the barrier passage problem of a
particle diffusing in a 2D non-orthogonal PES is studied in the
framework of statistical Langevin reactive dynamics. In the whole
range of friction strength from Ohmic to non-Ohmic damping, the
optimal incident angle of the diffusing particle is found to be
greatly influenced by the non-orthogonality of the system tensors. A
type of potential valley returning behavior of the optimal path is
witnessed in the 2D non-Ohmic damping environment under the combined
influence of off-diagonal system tensors and the coupling between
the two degrees of freedom. The result of this work provides useful
information to the study of stochastic dynamical reactive processes
such as the fusion of massive nuclei and those in connection with
the synthesis of super-heavy elements.

\section * {Acknowledgements}
This work was supported by the Scientific Research Starting
Foundation of Qufu Normal University and the National Natural
Science Foundation of China under Grant No. 10847101.


\end{document}